\documentstyle[preprint,aps,prc]{revtex}
\newcommand{\pr}{\protect}
\newcommand{\nuc}[2]{\mbox{$^{#1}$#2}}
\newcommand{\bea}{\begin{eqnarray}}
\newcommand{\eea}{\end{eqnarray}}
\newcommand{\beq}{\begin{equation}}
\newcommand{\eeq}{\end{equation}}
\newcommand{\Ga}{\alpha} 
\newcommand{\Gb}{\beta}
\newcommand{\Gg}{\gamma}

\newcommand{\Gd}{\delta}
\newcommand{\GD}{\Delta}
\newcommand{\Ge}{\epsilon}
\newcommand{\Gt}{\theta}

\newcommand{\non}{\nonumber}
\newcommand{\bct}[3]{<\!#1|#2|#3\!>}
\begin{document}
\draft
\title{The  spectral function for finite nuclei in the local density
                    approximation}
\author{D.\ Van Neck, A.E.L.\ Dieperink}
\address{Kernfysisch Versneller Instituut, Zernikelaan 25,
9747 AA Groningen, The Netherlands}
\author{E.\ Moya de Guerra}
\address{Instituto de Estructura de la Materia, CSIC, Serrano 119,
28006 Madrid, Spain}
\maketitle
\begin{abstract}
 The spectral function for finite nuclei is
computed within the
framework of the Local Density Approximation, starting from
nuclear matter spectral functions obtained with a realistic
nucleon-nucleon interaction.  The spectral function is
decomposed into a single-particle part and a ''correlated'' part;
the latter is treated in the local density approximation.
 As an application momentum distributions, quasi-particle strengths
and overlap functions for valence hole states, and the light-cone
momentum distribution in finite nuclei are computed.
\end{abstract}

\section{Introduction}

All information on the single-particle structure of nuclei is contained
in the spectral function $S(k,E)$, where $k$ and $E$ are the momentum
and the removal energy of the nucleon, respectively.
This quantity plays a central role in a variety of high-energy
nuclear reactions, such as in (inclusive and exclusive)
quasi-free electron scattering and deep-inelastic lepton scattering
on nuclei. With presently available many-body
techniques (method of correlated basis functions)
it can be computed quite accurately for nuclear
matter starting from a realistic nucleon-nucleon interaction \cite{Ben}.
On the other hand, for finite nuclei with $A>4$
it is much more difficult to calculate the spectral function without
 making severe approximations.

In this paper we compute the spectral function
 for finite nuclei in the local density approximation (LDA).
To this end the spectral function is decomposed in terms of a
single-particle contribution and a correlation part.
 The former part which varies strongly with mass
number, is represented by the generalized mean field approximation;
the latter which is rather insensitive to the finite size of
nuclei is treated in the LDA. As an input the results for
the nuclear matter spectral function, computed as a function of the
density by Fabrocini et al.\ \cite{Sic,prep}, are used.
In order to get an idea about the sensitivity to the details of the
high-momentum components we also use as an input a simple
  parametrization of the momentum distribution in nuclear matter
 given by Baldo et al.\ \cite{Bal}.

Several applications are considered. First, by integrating over the
removal energy, momentum distributions for finite nuclei with closed
shells are computed.
Next we consider three different aspects of the spectral function:
(1) the quasi-particle strength for
valence hole states and the radial shape of the overlap functions
which have been measured in $(e,e'p)$ reactions
in several closed shell nuclei;
(2) the distribution of s.p.\ strength at large removal energies
where there is an important  enhancement of the high-momentum
components in the spectral function;
(3) finally  the light-cone momentum distribution, which plays a
central role in the convolution approach to the EMC effect in
deep-inelastic scattering on nuclei is computed.

This paper is organized as follows.
In section 2  the LDA to the spectral function is
described; section 3 deals with various applications; section 4
contains a summary and discussion.

\section{ The spectral function in the Local Density Approximation}

Usually the spectral function for a finite system is expressed in
terms of a (truncated) single-particle basis $S_{\Ga\Gb }(E)
 = <0|c_\alpha^\dagger \delta(E-H+E_0) c_\beta |0>        $.
Since we are interested in the effects induced by (short-range)
nucleon-nucleon correlations on the spectral function,
we wish to be independent of the choice and size of the s.p.\ basis
and use the coordinate representation.

For a finite nucleus the hole spectral function can be expressed
as
\beq
\label{speca}
S_{A}(\vec{R},\vec{r},E)=
\bct{0(A)}{c^\dagger (\vec{r}_1 ) \Gd(E-\hat{H}+E_{0(A)})
c(\vec{r}_2 )}{0(A)}
\eeq
with $\vec{R}=\frac{1}{2}(\vec{r}_1+\vec{r}_2)$ and
$\vec{r}=\vec{r}_1-\vec{r}_2 $.

The spectral function (\ref{speca}) is normalized to the number of
particles

\beq
\label{specnorm}
\int d\vec{R}\int_{-\Ge_{F(A)}}^{+\infty}dE\,S_A (\vec{R},\vec{r}=0,E)=A.
\eeq
We use the convention
$\Ge_{F(A)}=(E_{0(A+1)}-E_{0(A-1)})/2$
 for the Fermi energy in a finite system.

In order to make the connection to nuclear matter it is useful to
introduce the Fourier transform with respect to the relative distance
$\vec{r}$:
\beq
\label{eq3}
S_A (\vec{R},\vec{k},E)=\frac{1}{(2\pi)^3}\int d\vec{r}
\mbox{e}^{-i\vec{k}\cdot\vec{r}} S_A (\vec{R},\vec{r},E).
\eeq
The r.h.s. of eq.(\ref{eq3}) can be interpreted as the spectral
function for a nucleon
at position $\vec{R}$ with momentum $\vec{k}$ \cite{Sic,prep}.
In particular the momentum
distribution in the ground state of a nucleus with mass $A$ is given by
\beq
n_A (k)=\int d\vec{R} \int_{-\Ge_{F(A)}}^{+\infty}
dE\,S_A (\vec{R},\vec{k},E).
\eeq

For our purpose of making a LDA, it is necessary to decompose the
spectral function and momentum distribution
into a (generalized) {\em single-particle} part and a
{\em correlation} part.
The reason for this can be seen if one considers the
 global application of the LDA to the (special) case of the
momentum distribution
\beq
\label{naive}
 n_A^{\rm GLDA} (k)=\frac{1}{2\pi^3}\int d\vec{R}\,\nu(\rho(R),k),
\eeq
where  $\nu(\rho,k)$ is the momentum density
for nuclear matter at density $\rho$ and with Fermi
momentum $k_F(\rho) = (\frac{3}{2}\pi^2\rho)^{1/3}$.
As noted in \cite{Str}, eq.(\ref{naive})
 is divergent in the limit $k\rightarrow 0$.
This problem can be circumvented by decomposing
 the momentum
distribution into a {\em mean-field} part  and  a {\em correlation}
part generated by dynamical excitations.
While the former part  can be calculated
using modern mean-field theories such as density dependent
Hartree-Fock,  the latter part can then be approximated by
applying the LDA to the correlation
part of $\nu(\rho,k)$ only, i.e.
\beq
\label{conv}
n_A (k)=n^{\rm MF}_A (k)+
\frac{1}{2\pi^3}\int d\vec{R}\,\Gd\nu_< (\rho(R),k)+
\frac{1}{2\pi^3}\int d\vec{R}\,\Gd\nu_> (\rho(R),k),
\eeq
with
\bea
n_A^{\rm MF}(k)&=&\frac{1}{(2\pi)^3}\int d\vec{R}
d\vec{r} \mbox{e}^{-i\vec{k}\cdot\vec{r}}
     n_A^{\rm MF}(\vec{R},\vec{r}),\non\\
\Gd\nu_< (\rho,k)&=&(\nu(\rho,k)-1)\Gt(k_F -k)\non\\
\Gd\nu_> (\rho,k)&=&\nu(\rho,k)\Gt(k -k_F),
\eea
and the mean-field density given by the sum over the occupied orbitals
\beq
\label{densmf}
n_A^{\rm MF}(\vec{R},\vec{r})= \sum_{h=1}^A n_{h}^{\rm MF}
(\vec{R},\vec{r})=\sum_{h=1}^A \phi_{h}^* (\vec{r}_1)\phi_h (\vec{r}_2).
\eeq

A similar divergence problem, which occurs if the LDA is applied to
the full spectral function, can be treated in an analogous way.
 In this case there are several ways
to decompose $S$ in different parts.
Let us first consider only the
mean-field part of the interaction felt by a nucleon in the nucleus,
in which case the spectral function has the simple form:
\beq
\label{specmf}
S_A^{\rm sp} (\vec{R},\vec{r},E)=
\sum_{h=1}^{A} n_{h}^{\rm MF}(\vec{R},\vec{r})
\Gd(E+\Ge_h),
\eeq
In addition to the hole orbitals appearing in
eqs.(\ref{densmf}-\ref{specmf}), the mean-field
also defines the normally empty particle orbitals $\phi_{p}$.
As for deviations from the mean-field approximation
it is convenient to distinguish the following two
 dynamical effects of the residual interaction:

\begin{enumerate}
\item Depletion and fragmentation of the hole strength.

Nucleon-nucleon correlations lead to both depletion and fragmentation
of the single hole strength in eq.(\ref{specmf}).
We assume that this depletion
and fragmentation of the strength related to the hole orbital $h$
 only depend on the single hole energy, $\Ge_h$; they
  are taken equal to the depletion and
fragmentation of the spectral function
$S_{\rm nm}(\rho,k,E)$ in nuclear matter at the local density.
Consequently the $\Gd$-function in the single-particle
part of the correlated spectral
function in eq.(\ref{specmf}) is replaced by the distribution function
\beq
\delta(E+\Ge_h) \rightarrow
S_{\rm nm}(\rho(R),k(\Ge_h),E).
\eeq
This ensures a realistic behaviour of the strength distribution at large
removal energies; it is well known that Lorentzian distributions become
inadequate away from the quasi-particle peak and cannot be used for
 the evaluation of e.g.\ mean removal energies.
The momentum $k(\Ge_h )<k_F$ corresponding to $\Ge_h$ is chosen
in such a way  that the s.p.\ energy  (relative to
the Fermi energy) $\Ge_{F(A)}-\Ge_h$
corresponds to the position of the quasi-hole peak in
the nuclear matter spectral function.
The fact that the finite volume of the
nucleus imposes discrete energies and bound
single-particle wave functions
$\phi_h$ (of well-defined angular momentum) is taken into account
by keeping the mean-field density $n_{h}^{\rm MF}(\vec{R},\vec{r})$
in eq.(\ref{specmf}).
The single-particle part is dominated by
the mean field, and the $k$-dependence is mainly determined by the
hole orbitals.
\item Partial occupancy of the normally empty states.

The scattering of nucleon-nucleon pairs into states above the Fermi
momentum leads to both a depletion of the occupied Fermi sea and to the
 presence of high-momentum components in the spectral function at large
 removal energies. Since this effect is expected to depend  only
weakly on the  shape of the orbitals $\phi_p$ in the finite nucleus
  the correlation part of $S_A$  is taken equal to
 the spectral function in nuclear matter for $k>k_F$ at
the local density $\rho(R)$:
\beq
\label{speccor}
S^{\rm cor}_A (\vec{R},\vec{k},E)
=\frac{1}{2\pi^3}S_{\rm nm}(\rho(R),k,E)\Gt(k-k_F (R)).
\eeq
We emphasize again that this  prescription cannot be extended to the
whole spectral function ($k<k_F$) since an integration over $\vec{R}$
would lead to a singular result for the momentum distribution at
$k=0$ (see eq.(\ref{naive})). The separation of the spectral function
into a single-particle part and a correlation part is thus quite natural,
as the low-momentum part of the spectral function depends crucially on
the finiteness of the system.
\end{enumerate}

In summary, we use
the following decompositon of the hole spectral function:
\beq
\label{specex}
S_A (\vec{R},\vec{k},E)
=S_A^{\rm sp}(\vec{R},\vec{k},E)+S_A^{\rm cor} (\vec{R},\vec{k},E)
\eeq
with
\beq
S_A^{\rm sp} (\vec{R},\vec{k},E)=
\sum_{h=1}^{A} n_{h}^{\rm MF}(\vec{R},\vec{k})
S_{\rm nm}(\rho(R),k(\Ge_h),E)
\eeq
and $S_A^{\rm cor} (\vec{R},\vec{k},E)$ given by eq.(\ref{speccor}).
\section{Applications}
\subsection{Nuclear matter input}
For the nuclear matter spectral functions we used the results obtained
by Fabrocini et al. \cite{Sic,prep} for densities
$\rho/\rho_0 =$ 0.25, 0.5, 0.75 and 1.0, with
$\rho_0=0.16\,\mbox{fm}^{-3}$ the nuclear matter density at equilibrium.
The spectral functions were calculated with correlated basis function
 (CBF) theory using the Urbana $v_{14}$ NN interaction with
addition of a three-nucleon force \cite{Lag}. The calculations involve
a variational determination of the correlated groundstate and of the 1h
and 2h1p states. Furthermore perturbative corrections (2p2h) to the
groundstate and (2h1p) to the 1h states are added.

As noted in ref.\cite{prep}
 a peculiar feature of the nuclear matter results is that
the momentum density does not approach the non-interacting one in the
limit of zero density. For example, one finds a rather constant
depletion, e.g. $\nu(\rho,k \approx 0)$
= 0.86, 0.87, 0.86 for $\rho= \rho_0, \frac{3}{4}\rho_0,$ and $
\frac{1}{2} \rho_0 $, respectively,
and similar results for the discontinuity $Z(\rho)$ of $\nu(\rho,k)$ at
 the Fermi surface. This behavior can be attributed to the fact that in
the variational approach the attraction of the NN interaction leads
to bound pairs and thus to clustering for low densities. One may
argue that this effect is an artefact of the variational approach
and is not a realistic density dependence to be used in the LDA for
finite nuclei.
Therefore we assumed that $\delta \nu(\rho,k) \rightarrow 0$
for $\rho \rightarrow 0$
 by a smooth extrapolation to zero for
densities lower than $\frac{1}{4} \rho_0$.
One should note that the final results are quite insensitive to this
procedure because the weight of small $\rho$ values is quite small.

To investigate the sensitivity to the nuclear matter
input, it is
 of interest to compare the results with those for
 simple parametrizations of the nuclear matter momentum
distribution as  a function of the density. Here we
 use the parametrization recently given by Baldo et al. \cite{Bal}:
\beq
\label{Bal}
\delta \nu(\rho,k) = \left\{
\begin{array}{l}
      -0.21 -0.13 \kappa -0.19 (1-\kappa) \ln (1-\kappa),  \
                       \kappa<1    \\
   0.21+ 0.3 \arctan x +0.82x \ln x,  \
                         1< \kappa <2,\; k<2 {\rm fm}^{-1}   \\
         \frac{k_{\rm F}^5}{7.0} \exp{(-1.6k)} ,  \ 2< k< 4.5
{\rm fm}^{-1}
 \end{array}
\right.
\eeq
 where $x=(\kappa -1)/(\kappa^2+1) $  and $\kappa =k/k_{\rm F}$.
This is an approximate parametrization valid for $k_{\rm F} >1
{\rm fm}^{-1} $, in the sense that for
$k<2 {\rm fm}^{-1}$ the result depends on $\kappa$ only.
Since we require that  $\delta \nu(\rho,k) \rightarrow 0 $
for lower densities, we  multiplied the above expression
for $\delta \nu(\rho,k)$ with $k_F^2$ for
$k_{\rm F} <1 {\rm fm}^{-1}$. We also properly renormalized the
momentum distribution for all densities considered.
The parametrization of eq.(\ref{Bal}) provided a good fit to nuclear
matter momentum distributions with $1<k_F<1.75\,\mbox{fm}^{-1}$
calculated to second order in the Brueckner reaction matrix
and using a separable representation of the Paris interaction.

In Figure~1 we compare $\nu(k)$ from the two prescriptions for
nuclear matter at two densities. Note that the former has a higher
depletion and somewhat stronger correlations.

 This is also reflected in Table~1, where we list
the kinetic energies  obtained with the spectral functions of \cite{Sic}
 and with
the parametrization in \cite{Bal}, as a function of the Fermi momentum
in nuclear matter.
For the kinetic energies obtained with the spectral functions of
\cite{Sic} we assumed an exponential extrapolation of the momentum
distribution for large momenta.
We note that the total kinetic energy in the interacting system
is roughly twice the
free value given by $T_{\rm free}=\frac{3k_F^2}{10m}$.

\subsection{Momentum distributions in finite nuclei}
The momentum distributions  for finite nuclei are computed
by integrating eq.(\ref{specex}) over $R$ and $E$
\beq
\label{mom}
n_A(k)=n^{\rm MF}_A(k)+\int d\vec{R}\sum_h n^{\rm MF}_h (\vec{R},\vec{k})
\Gd\nu_< (\rho(R),k(\Ge_h))
+\frac{1}{2\pi^3}\int d\vec{R}\Gd\nu_> (\rho(R),k),
\eeq
for the spherical closed-shell nuclei with $A=$16, 40, 48, 90 and 208.
We checked that the normalization (\ref{specnorm}) was in all cases
fulfilled to better
than 1\%.
The mean-field densities were taken from \cite{Cas}.
For comparison we also used the more conventional
form of the LDA of eq.(\ref{conv})  for the momentum density.
The two expressions (eq.(\ref{conv}) and eq.(\ref{mom})) differ
in the second term, which represents the depletion of the mean-field
momentum distribution at small momenta.
In Figure~2 the resulting momentum densities are compared for
\nuc{16}{O} and \nuc{208}{Pb}. As expected, the high-momentum
components ($k>1.6\,\mbox{fm}^{-1}$) are identical for both cases.
In the intermediate range ($1\,\mbox{fm}^{-1}<k<1.6\,\mbox{fm}^{-1}$)
the conventional LDA gives rise to a kink in the
momentum distribution. The kink is due to the second term in
eq.(\ref{conv}), which (with $k_{F_{\mbox{\scriptsize max}}}$ the Fermi
momentum
corresponding to the largest density in the nucleus) has a very steep
behaviour for
$k\,\,^{<}_{\rightarrow}\,\,k_{F_{\mbox{\scriptsize max}}}$
and is
strictly zero for $k>k_{F_{\mbox{\scriptsize max}}}$. This unphysical
feature is removed in the present approach, since the second term in
eq.(\ref{mom}) has a smooth behaviour for all $k$. The
present treatment also predicts for \nuc{16}{O} somewhat larger
$n_A (k)$ for $k\approx 0$ than the conventional LDA, because
eq.(\ref{mom}) takes into account that the reduction of the
mean-field distribution for $k\approx 0$ comes only from the
depletion of the 1s1/2 ($l=0$) orbitals, which are deeply bound and
therefore less depleted than the $1p$ orbitals. The conventional LDA,
on the other hand, can only take into account an average depletion.
  For \nuc{16}{O} we also included in Figure~2 the result of a
Variational Monte-Carlo calculation by Pieper et al. \cite{Piep}.
As was noted in \cite{prep}, there is good agreement up to
$k=2\,\mbox{fm}^{-1}$ between the LDA and the VMC approach.
For $k>2\,\mbox{fm}^{-1}$ the difference can probably be ascribed to
the fact that the VMC calculation was carried
out using the Argonne NN interaction, which has a stronger tensor
force than the Urbana interaction \cite{Piep}.

In Figure~3  the LDA result for the $A$-dependence of the momentum
density is shown.
Although with increasing $A$ there is a clear tendency towards the
nuclear matter momentum density, finite-size effects are seen to
remain important (even for \nuc{208}{Pb}), especially at small $k$.

The resulting $A$-dependence of the kinetic energy $T$,
$\Delta T = T-T_{\rm MF}$,
and the mean removal energy $<E>$ is given in Table~2.
It is seen that both the MF and the correlation part of the kinetic
energy amount to
approximately 17 MeV, and increase with $A$. The difference between
$\GD T$ in
the present approach and in the conventional LDA is small (at most 3\%),
whereas
the parametrization of the nuclear matter momentum density by Baldo et
al.\ leads also in finite nuclei to sizeably larger $\GD T$.
The mean removal energies $<E>$ in Table~2 were calculated using the
mean-field (\ref{specmf}) and correlated spectral
function (\ref{specex}), and have been corrected for the
mean-field rearrangement energy.
The inclusion of nucleon-nucleon correlations increases $<E>$,
from 33 MeV in the mean-field approximation
to about 55 MeV. Note that the effects of the Coulomb
interaction and the proton-neutron asymmetry are only taken into account
via the  mean-field single-particle energies.
For completeness we have also
compared in Table~2 the experimental binding energies per particle
$E_A /A$ with the binding energies following from the Koltun sum rule
\beq
E_A /A = \frac{1}{2}\left(\frac{A-2}{A-1}T-<E>\right),
\eeq
which is valid if only two-body forces are present. We find about
2.4 MeV too much binding. This discrepancy is not too worrisome since we
neglected the effect of three-body forces, and (part of) the Coulomb
interaction and proton-neutron asymmetry effects.
\subsection{Spectroscopic factor and radial shape of the quasi-hole
wavefunctions}
\subsubsection{The LDA approach}
Of special interest  is the spectral function for the least
bound orbitals just below the Fermi energy.
For values of the removal energy near the Fermi energy $\Ge_F$
the nuclear matter spectral function vanishes except for momenta
slightly below  $k_F$. As a consequence
the correlation part $S_A^{\rm cor}$ in eq.(\ref{speccor}) vanishes and
the spectral function in LDA can be expressed as
\beq
\label{limit}
S_A (\vec{R},E\approx \Ge_{F(A)})=\sum_{h}
\phi_{h}^* (\vec{R})\phi_h (\vec{R})
Z(\rho(R))\Gd(E+\Ge_h).
\eeq
By comparing eq.(\ref{limit}) with the general form of the spectral
function for the top shells in the $(A-1)$ nucleus
$$ S_A(\vec{R},E)=
\sum_h |<\psi_h^{\rm A-1}|c^\dagger(\vec{R})|\psi_A>|^2
   \delta(E+\Ge_h) =
\sum_h | \psi_h(\vec{R})|^2 \delta(E+\Ge_h) $$
one sees that in the LDA
the overlap functions $\psi_h(R) $ for the low-lying quasi-particle
states can be expressed as
\beq
\label{overlap}
\psi^{LDA}_h (\vec{R})=\sqrt{Z(\rho(R))}\phi_h (\vec{R}).
\eeq
This result (\ref{overlap}) has
the correct asymptotic behaviour of the overlap function which is
determined by the separation energy of each orbital \cite{vneck}.
It has also been noted
 that for the  case of droplets of \nuc{3}{He} atoms
eq.(\ref{overlap}) indeed provides a good approximation to
the exact overlap functions \cite{Lew}.

Recent $(e,e'p)$ and $(\Gg,p)$ experiments
(after correction for MEC) have probed the
spectral function for the top shells at large missing momenta.
Therefore it is of interest to examine to what extent
 the overlap functions for valence hole states are affected by
 high-momentum components due to short-range
NN interactions. One sees from eq.(\ref{overlap}) that in the LDA
 the only modification of the
mean-field wavefunction in coordinate space can be expressed in terms of
a modulation with a density dependent factor that reduces the
s.p.\ strength.

As a typical example Figure~4 shows that the net effect
on the radial shape of the $1p_{1/2}$ overlap function in \nuc{16}{O}
is a slight depletion of the interior and an enhancement of the
surface and tail region.
In momentum space this leads to only small global deviations from a
typical mean-field wavefunction.
This is in agreement with results obtained by M\"{u}ther and
Dickhoff \cite{Mut}, who computed the overlap function for the $1p_{1/2}$
orbital in $^{16}$O using Green function perturbation theory with a
G-matrix interaction derived from a realistic NN interaction.

The spectroscopic strength of a quasi-particle state $h$ is in this
approach given by:
\beq
Z^{LDA} = \int d\vec{R} Z(\rho(R)) |\phi_h (\vec{R})|^2.
\eeq
{}From  Figure~5 one sees that in the LDA the spectroscopic factors
(averaged over the valence hole states) decrease slowly with increasing
$A$. This is related to the increase of the mean density with $A$.
In comparison the data of ref.\cite{Nik} also show little
$A$-dependence, but
the experimental spectroscopic factors are about 0.1 smaller than the
LDA prediction.

\subsubsection{Inclusion of surface effects}

One effect not taken into account in a LDA treatment of
correlations is the occurrence of surface degrees of freedom in finite
nuclei. In particular collective
low-lying excitations in the core nucleus (surface vibrations) are known
to couple strongly to the valence s.p.\ states, and
may affect the shape of the radial overlap function.
Namely the valence hole state in the residual (A-1)-nucleus has
one-phonon/one-hole components which can have a non-vanishing
single-particle overlap with two-phonon components in the ground state of
the target nucleus. This leads to admixtures in the overlap function
of s.p.\ states which, in contrast to the hole mean-field wave functions,
do have sizable components in the momentum range
$1.5<k<2.5\,\mbox{fm}^{-1}$.

Following refs.\cite{Mahaux,Ma} we have included
 the effect of surface vibrations
on the overlap functions in \nuc{208}{Pb} through the energy dependence
of the mean field. In general both the non-locality and the genuine
energy dependence can be included by the introduction of a total
effective mass $m^* (R)$ as the product of a $k$-mass ($m_k$) and an
$E$-mass ($m_E$)
\beq
\frac{m^* (R)}{m}=\frac{m_k (R)}{m}\frac{m_E (R)}{m}.
\eeq
Since we assume that the non-locality (giving rise to
$m_k$) is already included in the Skyrme HF wavefunctions and that
all volume contributions to $m_E$
are already incorporated in the LDA,
we only take into account explicitly the surface peaked $E$-mass
$m_E (R)/m=(1+\Gb_S dg/dR)$. Here the central part of the
mean-field potential $g(R)$ and the parameter $\Gb_S$, which reflects
the coupling strength of the surface modes to the s.p. states,
were taken from \cite{Ma}.
We find that he resulting overlap function, $\psi_h^{SV} (\vec{R})$,
which includes the effect of surface vibrations, can in good
approximation be expressed by an extra modulation factor
in eq.(\ref{overlap})
\beq
\label{surface}
\psi_h^{SV} (\vec{R})\approx\sqrt{\frac{m_E (R)}{m}}
\sqrt{Z(\rho(R))}\phi_h (\vec{R}).
\eeq
We have checked e.g.\ that for the valence proton s.p.\ states
in \nuc{208}{Pb} the quasi-particle
wavefunctions in \cite{Ma} are well approximated by the standard
Woods-Saxon wavefunctions multiplied by the effective mass
according to (\ref{surface}), although the surface peaking is
somewhat overestimated. We note that this prescription was also
proposed in ref.\cite{prepnik}.

Figure~6 shows that in momentum space the momentum distributions
generated by eq.(\ref{surface}) are enhanced
compared to overlap functions without surface effects,
by a factor of 10-100 for momenta in the range
$1.5<k<2.5\,\mbox{fm}^{-1}$.
There is indeed recent evidence from an $(e,e'p)$
experiment on \nuc{208}{Pb} \cite{prepnik}
for a  systematic enhancement of the momentum distribution in this
range, as compared to  mean-field valence hole wavefunctions.
However, in a detailed analysis of the experimental data it should
be taken into account that the inclusion of surface effects through
a surface-peaked modulation factor changes e.g.\ the r.m.s. radius
of the overlap function, as well as
the momentum distribution at smaller momenta.

The extra reduction $Z^{SV}$ of the quasi-particle strength
that results from the coupling to surface vibrations
can be calculated from the expectation value of $(m_E (R)/m)^{-1}$
with respect to the overlap function $\psi_h^{SV}$.
For the total quasi-particle strength $Z=Z^{SV}.Z^{LDA}$
we obtain an average of $Z=0.50$ for the valence hole shells in
\nuc{208}{Pb}. This result is similar to the one obtained
in \cite{Benh}, but is 0.15 lower than the data in \cite{Nik}.
This is probably due to double counting when adding the surface effects
 to the nuclear matter results.
\subsection{Global distribution of single-particle strength}
The energy distribution of total proton hole strength
\beq
S_A (E)=\int d\vec{k} S_A (k,E)
\eeq
is shown in Figure~7 for \nuc{208}{Pb}, split in s.p.\ and correlation
parts. The s.p.\ part is dominant up to 50 MeV removal energy. It
represents the quasi-hole strength corresponding to knockout from
the various proton hole shells. Beyond the energy region of
the quasi-hole excitations the s.p.\ part of the strength dies out
quickly, and the correlation part of the strength, which extends
to very high removal energies, becomes dominant.

Also of interest is the difference in the momentum distribution
between small and large values of the removal energy.
Figure~8 clearly illustrates that high-momentum components
are correlated with large removal energies.
Qualitatively
this is easily understood, since in order to remove a nucleon with
high-momentum in the groundstate one has to break a correlated pair.
The remaining nucleon in the pair has roughly opposite momentum, its
kinetic energy leading to high excitation energies in the residual
system.
\subsection{Light-cone momentum distribution}
The light-cone momentum distribution plays a central role
in the description of high-energy reactions on nuclear targets,
such as deep-inelastic scattering.
For example, in the simple convolution approach
(use of the impulse approximation, and neglecting off-shell effects)
to the EMC effect
the structure function $F_2^A(x)$ for the $A$-body target can be
expressed as  \cite{Kul}
$$ F_2^A(x)= \int_x^\infty dy f_A(y) F_2^N(x/y) $$ where
the nuclear structure
information is contained in the light-cone momentum distribution
\begin{equation}
 f_A(y)=  \int dEd^3k S_A(k,E)y \delta(y- \frac{k^0 +k^3}{m}),
\end{equation}
where the factor $y$ represents the flux factor.
For values of $x< 0.6$ one does not need the full spectral function
$S_A(k,E),$ and it is
sufficient to expand $F^N_2(x/y)$ around $y=1$
\beq F_2(x/y)=F_2(x)-(y-1)xF_2^{'}(x)+\frac{1}{2}(y-1)^2(2xF_2^{'}(x)
    +x^2F_2^{''}(x)),
\eeq
and therefore only the lowest moments of $f(y)$ are needed.
The lowest moments of $f(y)$ (up to order $\frac{\epsilon^2}{m^2}$)
can  easily be expressed in terms of $T$ and $<E>$
$$ <y-1>= \frac{<E>+\frac{2}{3}T}{m}, \
    <(1-y)^2> = \frac{2T}{3m}.   $$
with the result
$$ F_2^A(x)/A= F_2^N(x) -\frac{<E>}{M}x F_2^{N'}(x)+\frac{T}{3M}x^2
    F_2^{N''}(x). $$
In the earlier analysis of the EMC effect for finite nuclei mostly
the MF result has been used for $<E>$ and $T$ \cite{Bic}, which
can explain only $30\% $ of the observed reduction at intermediate $x$.
The  inclusion of correlations enhances the effect
 in the ratio $R_A(x)= F_2^A(x)/AF_2^N(x)$ by about a factor 2,
 but still cannot fully explain the observed  EMC ratio
as a function of $x$ \cite{Die}.

 It is also of interest to examine whether the mass dependence
 predicted by the convolution method agrees with the observed one.
To this  end it is convenient to parametrize the $A$ dependence of
$T$ and $<E>$ in terms of a `volume' and a `surface' term, i.e.
$T=T_{\rm vol} -T_{\rm surf} A^{-1/3},$ and similarly for $<E>$, hence
$R_A(x)= c_1(x) +c_2(x) A^{-1/3}$.
As an example, using the best fit values to the LDA results for
$T_{\rm vol}$ (=39.38MeV) and $T_{\rm surf}$ (=18.87MeV),
and using the Koltun sum rule for the mean removal energies,
  one finds for $x=0.60$ that $R_A(x)= 0.83 + 0.18 A^{-1/3}$.

Recently a new analysis  of the $A$-dependence of deep-inelastic
electron scattering at SLAC was presented \cite{Arn}.
In that analysis the  EMC ratio for the range of $A$ considered
 ($2<A<197$) was  parametrized
as a power law: $R_A(x)=C(x)A^{\alpha(x)}$ with $C(x) \approx 1.0$;
e.g. at $x=0.60$  the best fit is obtained for $\alpha(x)=
-0.0346\pm 0.002$.
We note that this parametrization
 has a rather unphysical limit for $A \rightarrow \infty $.
If we use the more physical parametrization above we obtain
$c_1= 0.82, c_2= 0.23$
 which gives an equally good fit to the data (see Figure~9) and
moreover is in good agreement with
our prediction.
 This  indicates that the observed $A$
 dependence of the EMC effect is consistent with a purely
single nucleon effect. On the other hand if two-body effects would be
 important
a more general mass dependence would be expected.
\section{Summary and conclusions}
In this paper we have proposed a method to obtain the single nucleon
spectral function for finite nuclei from that of nuclear matter
by applying the local density approximation.
Recently a similar method for the construction of  the spectral function
in finite nuclei was proposed \cite{Sic} (with the same
nuclear matter input) and applied to inclusive
electron scattering at large $q$. The decomposition of the
spectral function  in \cite{Sic} is different from ours;
 e.g.\ at the variational level the s.p.\
part only contains the
direct $g.s.\rightarrow 1h$ contributions, whereas the
$g.s.\rightarrow 2h1p$ background contributions are part of the
correlation part for both $k < k_F$ and $k > k_F$. On the other hand
in our approach the correlation part contains the full nuclear matter
spectral function only for $k > k_F$, whereas the background
contribution for $k < k_F$ is incorporated in the generalized
single-particle part. However, the resulting momentum distributions
 for \nuc{16}{O}
from \cite{Sic} and from the present work are almost identical.

A point of uncertainty in the LDA treatment of nuclear correlations
is the extrapolation of the nuclear matter spectral functions
in the limit of zero density, but this is expected
to be of little influence on the final results.

We find that correlations lead to an appreciable depletion of
quasi-particle strength $Z$ in nuclei; in $r$-space
 the valence overlap orbital is somewhat more reduced
in the interior  than in the exterior region,
but we find that the momentum distribution
of the valence orbit overlap is hardly affected.
This agrees with the conclusion in ref.\cite{Mut}.
However, surface effects
 may lead, for intermediate momenta, to a considerable enhancement
of the momentum distribution for valence hole states.
\subsubsection*{Acknowledgement}
We would like to thank I.Sick, O.Benhar and A. Fabrocini for making
available to us the results of their calculations of nuclear matter
spectral functions at various densities. \\
This work has been supported in part by DGICYT (Spain) under contract
PB92/0021-C02-01. This work is part of the research program of the
foundation for Fundamental Research of Matter (FOM), which is financially
supported by the Netherlands Organisation for Advancement of Pure
Research (NWO).

\begin{table}[table1]
\caption{Kinetic energy [MeV] per particle as a function of Fermi
momentum in nuclear matter, calculated with the
momentum density from Baldo et al. \pr\cite{Bal}, and from
Fabrocini et al. \pr\cite{Sic,prep}}
\begin{center}
\begin{tabular}{|r|l|l|l|}
 $k_{\rm F} ({\rm fm}^{-1})$ &  1.13 & 1.23 & 1.33   \\
\hline
$T_{\rm free}$    & 15.88 &  18.82 & 22.00      \\
Baldo         & 37.62 &  42.93 & 48.31      \\
Fabrocini       & 33.25 &  36.99 & 40.77      \\
\end{tabular}
\end{center}
\end{table}
\begin{table}[table2]
\caption{The A-dependence of kinetic and mean removal energies [MeV]
per particle. $T^{MF}$ and $<E>^{MF}$ refer to the mean-field
approximation.
$\GD T ^{a)b)}$ were calculated with the nuclear matter input from
Fabrocini et al.\pr\cite{Sic,prep}, using ($^{a)}$) the present model
for the spectral
function and ($^{b)}$) the conventional LDA. For $\GD T ^{c)}$ the
nuclear matter input from Baldo et al.\pr\cite{Bal} was used, in the
conventional LDA. The mean removal energy $<E>$ was calculated with
the present LDA for the spectral function and includes the mean-field
rearrangement energy.
$E_A /A$ is the binding energy per particle, calculated using
the Koltun sumrule with $<E>$ and $T ^{a)}$.}
\begin{center}
\begin{tabular}{|r|l|l|l|l|l|}
$A$          &16    &40   &48   &90         &208    \\
\hline
$T^{MF}$   &15.4  &16.5 &17.7 &17.9&18.6   \\
$\GD T ^{a)}$&16.3  &16.9 &17.0 &17.3&17.5   \\
$\GD T ^{b)}$&15.8  &16.5 &16.7 &17.0&17.3   \\
$\GD T ^{c)}$&18.5  &20.1 &20.7 &21.4&22.2   \\
$<E>^{MF} $&30.4  &33.2 &34.8 &35.1&34.2   \\
$<E>        $&50.4  &54.0 &55.9 &56.5&56.6   \\
$E_A /A$ &-10.6 &-11.0&-11.1&-11.0&-10.5  \\
$E_A /A$(exp.)   &-8.0  &-8.6 &-8.7 &-8.7&-7.9   \\
\end{tabular}
\end{center}
\end{table}
\begin{figure}[fig1]
\caption{Nuclear matter momentum distributions (normalized to
$\frac{4}{3}\pi k_F^3$) at densities $\rho / \rho_0$ = .5 and 1,
according to the spectral function of Fabrocini et al.\
\pr\cite{Sic,prep} (full line) and with the parametrization of
Baldo et al.\ (dashed line).}
\end{figure}
\begin{figure}[fig2]
\caption{Momentum density (normalized to unity) in \nuc{16}{O} and
\nuc{208}{Pb}. Short-dashed line: mean-field result. Full line: present
LDA treatment. Long-dashed line: conventional LDA. The dots in the
\nuc{16}{O} plot were taken from \pr\cite{Piep}.}
\end{figure}
\begin{figure}[fig3]
\caption{Momentum density (normalized to unity) for finite nuclei
(dashed lines) and nuclear matter at equilbrium density (full line),
calculated in the present LDA treatment, with the nuclear matter input
taken from \pr\cite{Sic,prep}.}
\end{figure}
\begin{figure}[fig4]
\caption{Overlap function (normalized to unity) for the
\nuc{16}{O}-$1p1/2$ state in $r$-space and $q$-space. Dashed line:
mean-field approximation. Full line: including correlations in the LDA
according to eq.(\pr\ref{overlap}).}
\end{figure}\
\begin{figure}[fig5]
\caption{Average spectroscopic factor for the valence hole states,
as a function of mass number.  The LDA
prediction of the present work (squares) is compared with data (circles)
taken from a compilation of $(e,e'p)$ measurements in \pr\cite{Nik}.}
\end{figure}
\begin{figure}[fig6]
\caption{Overlap function (normalized to unity) in $q$-space for
valence proton hole states in \nuc{208}{Pb}.
Short-dashed line: mean-field approximation.
Full line: including correlations in the LDA
according to eq.(\pr\ref{overlap}).
Full line with dots: including additional surface effects according to
eq.(\pr\ref{surface}).}
\end{figure}
\begin{figure}[fig7]
\caption{Proton spectral function for \nuc{208}{Pb}, integrated over
all momenta, as a function of removal energy. Short-dashed line:
single-particle part. Long-dashed line: correlation part. Full line:
total.}
\end{figure}
\begin{figure}[fig8]
\caption{Proton spectral function for \nuc{208}{Pb}, integrated over
various regions (0-50, 50-100, 100-150 and 150-200 MeV)
of removal energy, as a function of momentum.}
\end{figure}
\begin{figure}[fig9]
\caption{Mass dependence of the EMC ratio $R_A (x)$, for $x=0.6$. The
SLAC data and best power law fit (long-dashed line) were taken from
\pr\cite{Arn}. Short-dashed line: best fit with the
parametrization in terms of a volume and surface contribution
(see text). Full line: LDA prediction for this parametrization.}
\end{figure}
\end{document}